\documentclass[prb,twocolumn,superscriptaddress,showpacs,amsmath,amssymb]{revtex4}
\usepackage{graphicx}
\usepackage{dcolumn}
\usepackage{bm}
\usepackage{color}
\newcommand{\nn}{\nonumber}
\newcommand{\beq}{\begin{eqnarray}}
\newcommand{\eeq}{\end{eqnarray}}

\begin{document}

\title{Odd-frequency Cooper pair amplitude around a vortex core in 
a chiral $p$-wave superconductor in the quantum limit 
}

\author{Takeshi Daino}
%\email{mizushima@mp.okayama-u.ac.jp}
\affiliation{Department of Physics, Okayama University,
Okayama 700-8530, Japan}
\author{Masanori Ichioka}
%\email{oka@mp.okayama-u.ac.jp}
\affiliation{Department of Physics, Okayama University,
Okayama 700-8530, Japan}
\author{Takeshi Mizushima}
%\email{mizushima@mp.okayama-u.ac.jp}
\affiliation{Department of Physics, Okayama University,
Okayama 700-8530, Japan}
\author{Yukio Tanaka}
%\email{ytanaka@nuap.nagoya-u.ac.jp}
\affiliation{Department of Applied Physics, Nagoya University, Nagoya 464-8603, Japan}
\date{\today}

\begin{abstract}

%We here clarify the relation between an odd-frequency $s$-wave Cooper pair amplitude and low-energy bound states 
%around a vortex core of a chiral $p$-wave superconductor. To capture the quantum limiting behaviors and exactly deal with Majorana zero modes, we self-consistently solve the Bogoliubov-de Gennes equation. We demonstrate that the vortex bound states contain the multifaceted properties, such as the odd-frequency $s$-wave pair amplitude and Majorana zero mode in the quantum regime. The identity between the bound states and odd-frequency pairing also holds in finite energy bound states for single vortex winding anti-parallel to the chirality, but not for parallel vortex winding. We also discuss a vortex state with the double winding number. 

Solving the Bogoliubov-de Gennes equation, 
we study the spatial structure of odd-frequency $s$-wave Cooper 
pair amplitudes at each quantized energy level of vortex bound states 
in chiral $p$-wave superconductors. 
For zero energy Majorana states,   
the odd-frequency $s$-wave pair amplitude has the same spatial 
structure as that of the local density of states 
even in atomic length scale. 
This relation also holds in finite energy bound states  
for single vortex winding anti-parallel to the chirality, 
but not for parallel vortex winding. The double winding vortex case is also studied. 

\end{abstract}


\maketitle

%74.20.Pq	Electronic structure calculations (for methods of electronic structure calculations, see 71.15.-m)
%74.20.Rp	Pairing symmetries (other than s-wave)
%74.25.Ha	Magnetic properties including vortex structures and related phenomena (for vortices, magnetic bubbles, and magnetic domain structure, see 75.70.Kw)
%74.25.Op	Mixed states, critical fields, and surface sheaths
%74.45.+c	Proximity effects; Andreev reflection; SN and SNS junctions
%74.70.Pq	Ruthenates
%74.81.-g	Inhomogeneous superconductors and superconducting systems, including electronic inhomogeneities

%\kword{ 
%Vortex states, 
%Chiral $p$-wave pairing, 
%Odd-frequency pair amplitude, 
%Bogoliubov-de Gennes equation,  
%Majorana state} 

%%%%%%%%%%%%%%%%%%%%%%%%%%%% 
\section{Introduction}

Studies of spatial-variation in unconventional superconductors give 
valuable information to know characters of unconventional superconductivity. 
Among these studies, such as for 
ferromagnet/superconductor junctions,\cite{Bergeret,BergeretRMP,Eschring,yokoyama,ya07sfs,linder} 
normal metal/superconductor interfaces,\cite{TanakaPRL2007,TanakaPRL2007v1,
Asano07},  and vortex states,\cite{Yokoyama,Yokoyama2} 
it has been shown that their low energy quasiparticle states 
are tightly related to the induced odd-frequency Cooper pair 
amplitude.\cite{TanakaJPSJ2012}  
Nowadays, it is accepted that 
the Andreev bound state \cite{TanakaReview} and 
anomalous proximity effect \cite{Proximity}
are interpreted as the emergence of  
the odd-frequency Cooper pairs.\cite{TanakaJPSJ2012}

Since quasiparticles in superconducting states have to satisfy 
the anti-commutation relation, the pair potential must change its sign after a permutation of two quasiparticles is done. 
The minus sign comes from the combination of 
orbital components (odd or even parities), 
spin component (singlet or triplet) 
and Matsubara frequency (even or odd frequency). 
Conventional even-parity spin-singlet pairing 
and odd-parity spin-triplet pairing 
are classified to the even-frequency pairing. 
On the other hand, within the odd-frequency pairing, 
even-parity spin-triplet pairing \cite{Berezinskii} and 
odd-parity spin-singlet pairing \cite{Balatsky}
are also possible.
Here, we study odd-frequency pair amplitudes induced
around a vortex in even-frequency pairing 
superconductors.\cite{Yokoyama,TanakaJPSJ2012} 
When spin-flip mechanism is absent for quasiparticles, 
an odd-frequency spin-singlet odd-parity (OSO) pair amplitude 
appears around vortex of 
even-frequency spin-singlet even-parity (ESE) superconductors, 
or an odd-frequency spin-triplet even-parity (OTE) pair amplitude  
appears around vortex of 
even-frequency spin-triplet odd-parity (ETO) superconductors. 
We discuss the latter case in this work.\cite{Yokoyama}

Most of previous studies for odd-frequency pairing were 
based on quasiclassical Eilenberger theory 
which is valid in the limit 
$2 (k_{\rm F}\xi)^{-1} \!=\! \Delta/E_{\rm F} \!\ll\! 1$ 
for superconducting gap $\Delta$ and Fermi energy $E_{\rm F}$. 
$k_{\rm F}$ is Fermi wave number and $\xi$ is superconducting coherence length. 
When the limit $\Delta/E_{\rm F} \!\ll\! 1$ is not 
satisfied in the strong-coupling superconductors or 
we focus fine energy structures,
low-energy bound states around a vortex core are quantized to 
Caroli-de Gennes-Matricon (CdGM) states with energy splitting of 
the order $\Delta^2/E_{\rm F}$.~\cite{Caroli,Hayashi} 
The quantum nature associated with CdGM states has been directly observed through the STM-STS in an anisotropic superconductor~\cite{kaneko} and the quantum depletion of the particle density in Fermi gases.~\cite{MIT}
To study the quantum limit case within $\Delta \!\sim\! E_{\rm F}$, we have to consider the eigen-states 
by solving the Bogoliubov-de Gennes (BdG) equation, 
without using the quasiclassical approximation. 

Recently vortex bound states in chiral $p$-wave superconductors 
attract much attention, because Majorana state 
appears at exactly zero energy in the bound 
states.\cite{ReadGreen,ivanov,MizushimaPRL2008,MizushimaPRA2010,tewari,gurarie}  
In a Majorana zero mode of the half quantum vortex, particle and hole states are equivalent each other, giving rise to the non-Abelian statistics of its host vortices.~\cite{ivanov} 
Note that in the case of an integer vortex state with Majorana zero modes, the statistics is nontrivial.~\cite{kawakami,yasui}
On the other hand, the quasiclassical study 
in superfluid $^3{\rm He}$ and superconductors showed the relation between 
odd-frequency pair amplitudes 
and the local density of states (LDOS) in Majorana bound state.\cite{Higashitani,TanakaPRL2007}
Most recently, the issue on the relation between Majorana zero modes and odd frequency pairing has been addressed in a quantum nanowire.~\cite{asano2012}
The structures of bound states are distinguished by  
the direction of vortex winding, {\it i.e.}, parallel or anti-parallel 
to the chirality of chiral $p$-wave superconductivity.\cite{Matsumoto} Within the quasiclassical theory, 
it has been revealed that the zero energy DOS in the parallel (anti-parallel) vortex is fragile (robust)
against nonmagnetic impurities,~\cite{kato1,kato2} which is understandable with the odd-frequency pairing localized at the vortex.~\cite{tanuma}
Hence, since the zero energy state has many facets, it is important to study how the odd-frequency pair amplitude is related 
to the bound states, including Majorana state, in chiral $p$-wave superconductors. 

The purpose of this paper is to clarify the relation between the CdGM states and the odd-frequency pairing. To capture the quantum limiting behaviors inside the core, we here utilize the BdG equation which is capable of describing the rapid oscillations of wave functions in atomic length scale of $k_{\rm F}^{-1}$ and exactly deals with the Majorana mode. The studies in the quantum regime have not been done with the quasiclassical theory,~\cite{Yokoyama,Yokoyama2,tanuma} because of the lack of the $k^{-1}_{\rm F}$-scale physics. We also note that in previous works, such as Refs.~\onlinecite{Bergeret,BergeretRMP,Eschring,yokoyama,ya07sfs,linder,TanakaPRL2007,TanakaPRL2007v1,Higashitani}, odd-frequency pairings have been discussed in connection with surface bound states. However, the vortex bound state which we consider here may exhibit distinct behaviors from surface one in the sense that each level of the CdGM states is well isolated in the scale of $\Delta^2/E_{\rm F}$. In addition, the quasiparticle structure inside the core is determined by not only the bulk topology but also the vortex winding number.~\cite{MizushimaPRA2010,tewari,gurarie} Hence, we here unveil how the vortex winding number affects the odd-frequency pairing around the core.

%The purpose of this paper is to clarify how the vortex bound state 
%is related to the odd-frequency pairing, 
%comparing the odd-frequency $s$-wave pair amplitude and the LDOS 
%in each CdGM state.  
%Solving the BdG equation, we can study rapid oscillations of wave functions 
%in atomic length scale of $k_{\rm F}^{-1}$ and exactly deal with the zero energy state.
%In the quasiclassical theory, quantum oscillations of the order $k_{\rm F}^{-1}$ 
%are factorized out, and only the behaviors in the length scale of $\xi$ are studied. 

%While we studied the case of the $s$-wave pairing and 
%the chiral $p$-wave pairing, we mainly report the latter case here. 

%%%%%%%%%%%%%%%%%%%%%%%%%%%% 
\section{Formulation of Bogoliubov-de Gennes equation} 

We consider a cylindrical system of 
isotropic chiral $p_+$ wave superconductor with radius $R=200 k_{\rm F}^{-1}$ and two-dimensional isotropic Fermi surface.~\cite{MizushimaPRL2008,MizushimaPRA2010,Matsumoto}   
Therefore, the cylindrically symmetric order parameter with vortex winding $w \!\in\! \mathbb{Z}$ is given by
\begin{eqnarray}
\Delta({\bf r},{\bf k})
= \Delta_+(r) {\rm e}^{{\rm i}w \theta} Y_+({\bf k}) 
+ \Delta_-(r) {\rm e}^{{\rm i}(w+2) \theta}Y_-({\bf k}) , 
\label{eq:D}
\end{eqnarray} 
at the position ${\bf r}\!=\!r(\cos\theta,\sin\theta)$, 
where 
the pairing function is given by 
$Y_\pm({\bf k})=(k_x\pm{\rm i}k_y)/\sqrt{2}k_{\rm F} $ 
for relative momentum ${\bf k}=(k_x,k_y)$.  
In the $p_+$-wave superconductor, 
the Cooper pair has internal angular momentum $L_z=1$ along $z$-axis. 
In the right-hand side of Eq.~(\ref{eq:D}), 
the first term is the dominant $p_+$-component with a vortex, 
and the second term of $p_-$ wave indicates 
the small induced component of opposite chirality around the vortex. 

The eigen-energy $E_{\bf q}$ and wave functions 
$u_{\bf q}({\bf r}),v_{\bf q}({\bf r})$ of quasiparticles are calculated 
by the BdG equation~\cite{MizushimaPRL2008,MizushimaPRA2010,Matsumoto}    
\begin{eqnarray} 
\left(\begin{array}{cc} 
H_0({\bf r}) & \Pi({\bf r}) \\ 
-\Pi^\ast({\bf r}) & -H_0({\bf r}) \\  
\end{array}\right)
\left(\begin{array}{c} 
 u_{\bf q}({\bf r}) \\ v_{\bf q}({\bf r}) \\ 
\end{array}\right) 
=E_{\bf q} 
\left(\begin{array} {c}
 u_{\bf q}({\bf r}) \\ v_{\bf q}({\bf r}) \\ 
\end{array}\right) , 
\label{eq:BdG}
\end{eqnarray} 
where 
$H_0({\bf r})\!=\!(\partial_x^2 +  \partial_y^2 )/2M -E_{\rm F} $ 
with mass $M$, and 
$\Pi({\bf r}) \!=\!- \frac{{1}}{2}\sum _m\{ \mathcal{P}^{(m)}, 
\Delta_m ({\bf r}) \}$ with $m \!=\! \pm$ and 
$\mathcal{P}^{(\pm)} \!=\! \mp i{\rm e}^{\pm {\rm i}\theta}
(\partial_{r} \pm {\rm i}r^{-1} \partial_{{\theta}})/\sqrt{2}k_{\rm F}$ 
in chiral $p$-wave superconductors.  
$k_{\rm F}\!=\!(2ME_{\rm F})^{1/2}$. 
Throughout this letter, we use units $\hbar \!=\! k_{\rm B} \!=\! 1$.
For the cylindrical symmetric system, wave functions are given by  
\begin{eqnarray} 
u_{\bf q}({\bf r})=\tilde{u}_{\bf q}(r){\rm e}^{{\rm i}q_\theta \theta}, \quad 
v_{\bf q}({\bf r})=\tilde{v}_{\bf q}(r){\rm e}^{{\rm i}(q_\theta -w-1)\theta} . 
\end{eqnarray} 
The eigen-states are labeled by ${\bf q}\!=\!(q_\theta,\nu)$, 
where $\nu$ assigns eigen-states of the BdG equation for 
real functions $\tilde{u}_{\bf q}(r)$ and $\tilde{v}_{\bf q}(r)$. 
To numerically solve the BdG equation,
we utilize the Bessel function expansion method, where $\tilde{u}_{\bf q}(r)$ and $\tilde{v}_{\bf q}(r)$ are expanded in terms of the Bessel functions $J_{q}(r)$ as
\beq
\tilde{u}_{\bf q}(r) = \sum _j C_j \varphi^{(q_\theta)}_{j}(r), \hspace{2mm}
\tilde{v}_{\bf q}(r) = \sum _j D_j \varphi^{(q_\theta-w-1)}_{j}(r),
\eeq
where the normalization condition imposed on $u_{\bf q}({\bf r})$ and $v_{\bf q}({\bf r})$ reduces to $2\pi \sum _j ( C_j^2 + D_j^2 ) \!=\! 1$. Here, we introduce the set of the orthogonal functions, 
$\varphi^{(q)}_{j}(r) \!=\! \frac{\sqrt{2}}{R|J_{q+1}(k^{(q)}_{j}R)|}J_{q}(k^{(q)}_{j}r)$, where $k^{(q)}_{j} \!\equiv\! \frac{\alpha^{(q)}_j}{R}$ and $\alpha^{(q)}_j$ is the $j$-th zero of $J_q(r)$. This reduces the BdG equation (\ref{eq:BdG}) to a matrix eigenvalue problem.~\cite{Hayashi,Gygi,TanakaSSC}

For simplicity we assume 
that the pairing interaction $g_p$ works only for chiral $p$-wave components 
and that the quantization axis of spin is 
parallel to the ${\bm d}$-vector of spin-triplet pairing. 
This implies a singular vortex state, accompanied by spin-degenerate Majorana zero modes.~\cite{kawakami} Although in the half quantum vortex with a non-Abelian Majorana mode the ${\bm d}$-vector is transverse to the spin quantization axis, the low energy quasiparticles are commonly describable with the BdG equation (\ref{eq:BdG}).~\cite{ivanov}
Here we treat Zeeman energy to be negligible at enough low fields. 
Then, the selfconsistent condition $\Delta_{\pm}({\bf r})$ for 
chiral $p$-wave pair potentials is given with 
the imaginary part of the retarded Green's function $\cal{F}$ by 
\begin{eqnarray}
\Delta_{\pm}({\bf r}) = g_p \sum_{|E|<E_{\rm cut}} 
{\cal F}_{\pm,{\rm triplet}}({\bf r},E) f(E) ,
\label{eq:Dp1}
\end{eqnarray}
where ${\bf r}$ denotes the center-of-mass coordinate.
Here, $f(E)$ denotes the Fermi distribution function and 
even-frequency $p_\pm$-wave pair amplitudes are given by 
$ {\cal F}_{\pm,{\rm triplet}}({\bf r},E) 
\!=\!( {\cal F}_{\pm,\uparrow\downarrow}({\bf r},E)
+{\cal F}_{\pm,\downarrow\uparrow}({\bf r},E) )/2
$
with 
\begin{eqnarray} 
{\cal F}_{m,\downarrow\uparrow}({\bf r},E)
&=& \sum_{\bf q} \lim _{{\bf r}_{12}\rightarrow 0}
\mathcal{P}^{(m)\ast}_{12}\!
\left[ u_{\bf q}({\bf r}_1) v^{\ast}_{\bf q}({\bf r}_2)  
\right] \nn \\
&& \times \delta(E-E_{\bf q}), 
\label{eq:Dp4}
\end{eqnarray}
using $\mathcal{P}^{(m)}_{12}$ obtained from $\mathcal{P}^{(m)}$ 
with ${\bf r} \!\rightarrow\! {\bf r}_{12}={\bf r}_2-{\bf r}_1$. 
${\cal F}_{m,\uparrow\downarrow}$ is obtained from Eq.~(\ref{eq:Dp4}) with replacing $(u_{\bf q},v_{\bf q},E_{\bf q})$ to $(v^{\ast}_{\bf q},u^{\ast}_{\bf q},-E_{\bf q})$. 
It is found that for ETO superconductors without a magnetic Zeeman term, 
even-frequency spin-triplet $p_{\pm}$-wave pair amplitude 
${\cal F}_{\pm,\uparrow\downarrow}({\bf r},E)$ 
satisfies 
${\cal F}_{\pm,\uparrow\downarrow}({\bf r},E)
\!=\!-{\cal F}_{\pm,\uparrow\downarrow}({\bf r},-E)$ and 
${\cal F}_{\pm,\uparrow\downarrow}({\bf r},E)
\!=\!{\cal F}_{\pm,\downarrow\uparrow}({\bf r},E)$.~\cite{TanakaPRL2007}
The BdG equation (\ref{eq:BdG}) with Eqs.~(\ref{eq:Dp1}) and (\ref{eq:Dp4}) gives  
a closed set for the self-consistent calculation of $\Delta_\pm({\bf r})$ 
and wave functions of the eigen-states. 

From the selfconsistent solutions, 
we calculate the LDOS with $u_{\bf q} \!\equiv\! u_{\bf q}({\bf r})$ and $v_{\bf q} \!\equiv\! v_{\bf q}({\bf r})$ as
%$N({\bf r},E) \!=\! N_\uparrow({\bf r},E)+N_\downarrow({\bf r},E)$, where 
\begin{eqnarray} %&& 
N({\bf r},E) = \sum _{\bf q} \left[ \left|u_{\bf q}\right|^2 \delta(E-E_{\bf q})
+\left|v_{\bf q}\right|^2 \delta(E+E_{\bf q})\right].
\label{eq:N2}
\end{eqnarray}
The odd-frequency spin-triplet $s$-wave pair amplitude is given by 
${\cal F}_{s,{\rm triplet}} 
\!=\! ({\cal F}_{s,\uparrow\downarrow}
\!+\! {\cal F}_{s,\downarrow\uparrow})/2$ with 
\begin{eqnarray} &&
{\cal F}_{s,\uparrow\downarrow}({\bf r},E)
=  \sum_{\bf q} v^{\ast}_{\bf q}({\bf r}) u_{\bf q}({\bf r})  \delta(E-E_{\bf q})
\nonumber
\label{eq:Ds1}
\\ && 
{\cal F}_{s,\downarrow\uparrow}({\bf r},E)
=  \sum_{\bf q} u_{\bf q}({\bf r}) v^{\ast}_{\bf q}({\bf r}) \delta(E+E_{\bf q}).
\label{eq:Ds2}
\end{eqnarray}
Note that $e^{-i(w+1)\theta}{\cal F}_{s,\downarrow\uparrow}({\bf r},E)$ can be real, since $u_{\bf q}(r)$ and $v_{\bf q}(r)$ are obtained as a real function from Eq.~(\ref{eq:BdG}) with an appropriate ${\rm U}(1)$ phase of $\Delta _{\pm}$.   
Since ETO superconductors without a Zeeman field hold ${\cal F}_{s,\uparrow\downarrow}({\bf r},E)
\!=\! {\cal F}_{s,\uparrow\downarrow}({\bf r},-E)$ and ${\cal F}_{s,\uparrow\downarrow}({\bf r},E)
\!=\! {\cal F}_{s,\downarrow\uparrow}({\bf r},E)$, 
spin-singlet components in ETO superconductors, such as 
${\cal F}_{s,\uparrow\downarrow}({\bf r},E)
-{\cal F}_{s,\downarrow\uparrow}({\bf r},E)$, vanish.~\cite{TanakaPRL2007}
Equation~(\ref{eq:Ds2}) is scaled so that 
${\rm max} \{ e^{-i(w+1)\theta}{\cal F}_{s,{\rm triplet}}({\bf r},E=0) \}
\!=\! {\rm max} \{ N({\bf r},E=0) \}$.  
In figures, the LDOS and pair amplitude are normalized by these maximum values 
at $E \!\sim\! 0$. %Spin singlet components such as 
%${\cal F}_{s,\uparrow\downarrow}({\bf r},E)
%-{\cal F}_{s,\downarrow\uparrow}({\bf r},E)$ 
%vanish, since the Zeeman effect is not included. 

%%%%%%%%%%%%%%%%%%%%%%%%%%%% 
%\section{Order parameter} 
\section{Numerical results}

%%%%%%%%%%%%%%%%%%%%%%%%%%%%% 
\begin{figure}[tb!]
\includegraphics[width=8.0cm]{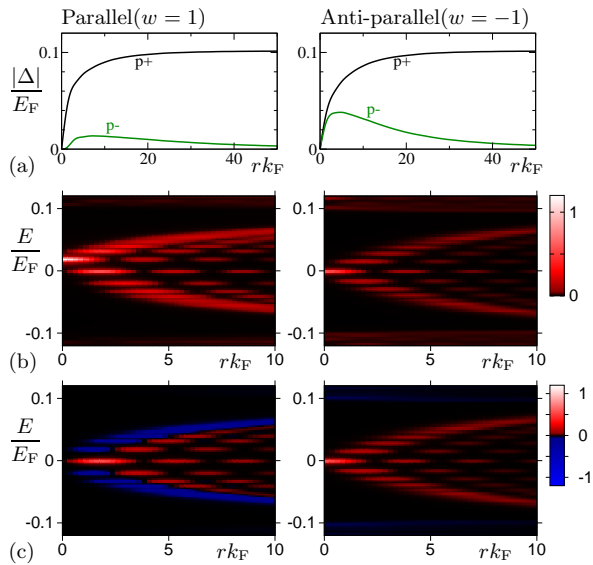}
\caption{
(Color online) 
(a) Profile of self-consistent chiral $p$-wave pair potential 
$|\Delta_\pm({\bf r})|$ around a vortex
as a function of radius $r$ from vortex center. 
Density plot of (b) $N({\bf r},E)$ and 
(c) ${\cal F}_{s,{\rm triplet}}({\bf r},E)$ with $\theta\!=\!0$. 
Left (right) panels are for the parallel (anti-parallel) vortex winding. 
}
\label{f1}
\end{figure}
%%%%%%%%%%%%%%%%%%%%%%%%%%%%% 

In our calculation, we set $g_p\!=\! 1.65$, 
$E_{\rm cut} \!=\! 2.0E_{\rm F}$, $T\!=\!0$, 
so that $\Delta \!\sim\! 0.1 E_{\rm F}$ when $w\!=\!0$.    
First, we consider two cases of 
the parallel ($w\!=\!1$) and anti-parallel ($w\!=\!-1$) vortex state. 
The obtained pair potentials $|\Delta_\pm({\bf r})|$ 
are shown in Fig.~\ref{f1}(a).  
Around the vortex core of $\Delta_+({\bf r})$, 
the opposite chiral component $\Delta_-({\bf r})$ is induced. 
The induced component for $w\!=\!1$ is smaller than that for $w\!=\!-1$, 
since the winding of $\Delta_-({\bf r})$, $w+2\!=\!3$, is larger than
$w+2\!=\!1$ in the anti-parallel case. 

%%%%%%%%%%%%%%%%%%%%%%%%%%%% 
%\section{Density plot of LDOS and odd-frequency pairing amplitude}

We examine the relation of LDOS $N({\bf r},E)$ and 
odd-frequency pair amplitude ${\cal F}_{s,{\rm triplet}}({\bf r},E)$ 
near the vortex core, which are shown in Figs.~\ref{f1}(b) and \ref{f1}(c) 
as a function of $r$ and $E$. 
There, we see the quantization of CdGM bound states on $E$, 
and the quantum oscillations of wave functions with respect to $r$. 
For parallel vortex winding, 
the LDOS $N({\bf r},E)$ is not symmetric for $E \!\leftrightarrow\! -E$, 
since the low energy LDOS at $r\!=\!0$ 
is finite only at positive energy as shown in the left panel of Fig.~\ref{f1}(b). 
On the other hand, the odd-frequency 
${\cal F}_{s,{\rm triplet}}({\bf r},E)$ is symmetric for $E \!\leftrightarrow\! -E$,~\cite{memo}
and shows oscillation between positive and negative values 
as a function of $r$, as shown in the left panel of Fig.~\ref{f1}(c).    
It is seen from the right panels of Figs.~\ref{f1}(b) and \ref{f1}(c) that 
for anti-parallel vortex winding, ${\cal F}_{s,{\rm triplet}}({\bf r},E)$ has the spectral evolution similar to $N({\bf r},E)$.  
Both are positive and symmetric for $E \!\leftrightarrow\! -E$ at $|E| \!<\! \Delta$.

%%%%%%%%%%%%%%%%%%%%%%%%%%%% 
%\section{LDOS and odd-frequency pairing amplitude in zero-energy Majorana state and 1st and 2nd excited bound states}

\begin{figure}
\includegraphics[width=8.0cm]{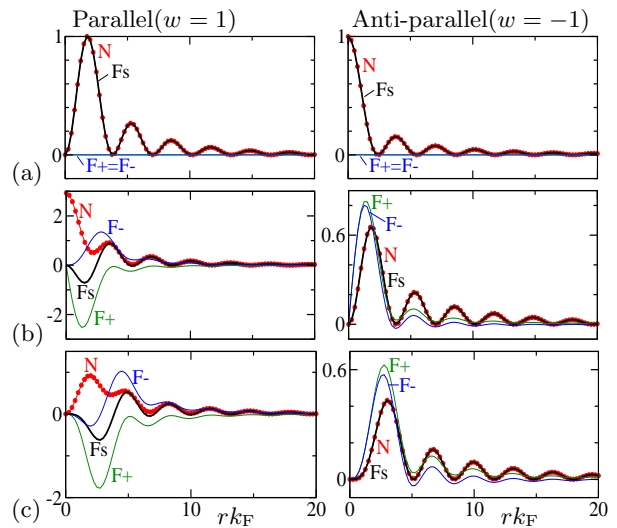}
\caption{
(Color online) 
$r$-dependence of LDOS $N({\bf r},E)$, ${\cal F}_{s,{\rm triplet}}({\bf r},E)$, and 
${\cal F}_{\pm,{\rm triplet}}({\bf r},E)$ 
at $\theta=0$; (a) Zero-energy Majorana bound state, and (b) the second and 
(c) third lowest energy bound states. 
Left (right) panels are for the parallel (anti-parallel) vortex winding. 
}
\label{f2}
\end{figure}

To examine whether the odd-frequency pair amplitude has 
the spatial structure exactly same as the LDOS or not, 
we compare the profiles as a function of $r$ for each energy level. 
Figure \ref{f2}(a) shows 
$N({\bf r},E)$ and 
${\cal F}_{s,{\rm triplet}}({\bf r},E)$, 
${\cal F}_{p,{\rm triplet}}({\bf r},E)$ 
for Majorana bound state at $E\!=\!0$.  
Here, we confirmed that the relations 
$N({\bf r},E=0)\!\propto\!{\cal F}_{s,{\rm triplet}}({\bf r},E=0)$
and 
\beq
{\cal F}_{\pm,{\rm triplet}}({\bf r},E=0) = 0, 
\label{eq:Feven}
\eeq 
at $\theta\!=\! 0$ in the fully quantum level of $k_{\rm F}^{-1}$, by solving the BdG equation (\ref{eq:BdG}).
This implies that the LDOS of the Majorana state consists of odd-frequency pairing amplitudes without including even-frequency ones. 
Substituting the Majorana condition $u_{\bf q}({\bf r}) \!=\! v_{\bf q}^*({\bf r})$ into Eqs.~(\ref{eq:N2}) and (\ref{eq:Ds2}) at $E \!=\! 0$,  
we obtain exactly 
\begin{eqnarray}
N({\bf r},E=0) \propto {\rm e}^{-i(w+1)\theta}{\cal F}_{s,{\rm triplet}}({\bf r},E=0).
\label{eq:Fodd}
\label{N-OFs}
\end{eqnarray} 
In the zero-energy state for $w \!=\! 1$, 
$N({\bf r},E=0) \!=\! {\cal F}_{s,{\rm triplet}}({\bf r},E=0) \!=\! 0$ 
at the vortex center $r \!=\! 0$, because 
both $u_{\bf q}({\bf r})$ with $q_\theta \!=\! 1$ and 
$v_{\bf q}({\bf r})$ with $q_\theta-w-1 \!=\! -1$ have non-zero winding.
For anti-parallel winding $w \!=\! -1$, 
since $q_\theta \!=\! 0$ both for $u$ and $v$, 
$N({\bf r},E=0)$ and ${\cal F}_{s,{\rm triplet}}({\bf r},E=0)$ 
have peak at $r=0$. 
The difference by the vortex winding direction is due to 
the angular momentum of CdGM vortex bound states.  

%%%%%%%%%%%%%%%%%%%%%%%%%%%%% 
%%%%%%%%%%%%%%%%%%%%%%%%%%%%% 

The relation of the LDOS and the odd-frequency pair amplitude was studied 
by the quasiclassical theory for surface bound states in superfluid 
${\rm ^3He}$ and superconductors.~\cite{Higashitani,TanakaPRL2007}
However, since the rapid oscillation of the length order $k_{\rm F}^{-1}$ 
is factorized out by the quasiclassical approximation, 
the theory discusses spatial variations only in the length scale of superconducting coherence length.
This corresponds to the behavior of envelop function of the oscillating behavior in Fig.~\ref{f2}. 
Note that the STM-STS with a $0.1$nm spatial resolution at low temperatures~\cite{kaneko} has succeeded in unveiling quantum limiting behaviors of a vortex core, whose microscopic structures are well understandable with the BdG theory but not with the quasiclassical theory.

Next, we study bound states at $E \!\ne\! 0$ 
in Figs.~\ref{f2}(b) and \ref{f2}(c).  
In the parallel vortex winding case (left panels), 
the second lowest energy state at $E \!=\! 0.019E_{\rm F}$ has 
$q_\theta \!=\! 0$ for $u$ and 
$q_\theta-w-1 \!=\! -2$ for $v$. 
Thus, $N({\bf r},E) \!\sim\! |u|^2$ has a sharp peak at the vortex center $r \!=\! 0$  
and the odd-frequency pair amplitude 
${\cal F}_{s,{\rm triplet}}({\bf r},E) 
\!\sim\! uv^\ast \propto {\rm e}^{{\rm i}2\theta}$ vanishes at $r\!=\!0$. 
In Figs.~\ref{f2}(b) and \ref{f2}(c), ${\cal F}_{s,{\rm triplet}}({\bf r},E)$ also 
shows oscillation between positive and negative values at $r \!>\! 0$, implying $N({\bf r},E) \!\ne\! {\cal F}_{s,{\rm triplet}}({\bf r},E)$. 
The third lowest energy state at $E\!=\!0.032E_{\rm F}$ has 
$q_\theta\!=\!-1$ for $u$ and $q_\theta-w-1\!=\!-3$ for $v$. 
Also in this state, we see that 
$N({\bf r},E) \!\ne\! {\cal F}_{s,{\rm triplet}}({\bf r},E)$.  
For the even-frequency pairing amplitude, we see that 
${\cal F}_{+,{\rm triplet}}({\bf r},E) \!\ne\! {\cal F}_{-,{\rm triplet}}({\bf r},E) $.

For the anti-parallel $w \!=\! -1$, as shown in the right panels of Figs.~\ref{f2}(b) and \ref{f2}(c), 
the second lowest energy state at $E\!=\!0.016E_{\rm F}$ has $q_\theta\!=\!1$ both 
for $u$ and $v$. 
The third lowest energy state at $E\!=\!0.029E_{\rm F}$ has $q_\theta\!=\!2$ 
for $u$ and $v$. 
Even in these cases, we find that 
$N({\bf r},E) \!\propto\! e^{-i(w+1)\theta}{\cal F}_{s,{\rm triplet}}({\bf r},E)$ approximately holds, that is, the relation in Eq.~(\ref{N-OFs}) can be extended to a finite $E$.
This comes from the symmetric LDOS structure for $E \leftrightarrow -E $ and the $\theta$-independent structure 
$uv^\ast \!\propto \! {\rm e}^0$ when $w\!=\!-1$, since
the BdG equation (\ref{eq:BdG}) holds the particle-hole symmetry, $(u,v) \!\leftrightarrow\! (v^\ast,u^\ast)$
for $E \!\leftrightarrow\! -E$. 
We also see that 
${\cal F}_{+,{\rm triplet}}({\bf r},E) \!\sim\! 
 {\cal F}_{-,{\rm triplet}}({\bf r},E) $, 
but there are small deviations between them.

%%%%%%%%%%%%%%%%%%%%%%%%%%%% %%%%%%%%%%%%%%%%%%%%%%%%%%%% 
%\section{$E$-dependence of LDOS and odd-frequency pairing amplitude}

%%%%%%%%%%%%%%%%%%%%%%%%%%%%% 
\begin{figure}[t!]
\includegraphics[width=8.0cm]{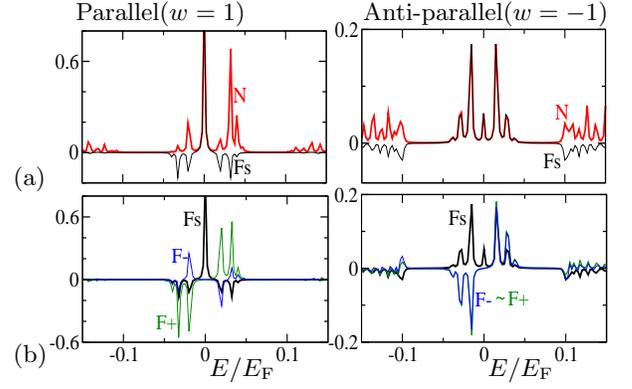}
\caption{
(Color online) 
(a) $E$-dependence of $N({\bf r},E)$ and ${\cal F}_{s,{\rm triplet}}({\bf r},E)$ 
at $r=2 k_{\rm F}^{-1}$ and $\theta=0$. 
(b) $E$-dependence of ${\cal F}_{s,{\rm triplet}}({\bf r},E)$ and 
${\cal F}_{\pm,{\rm triplet}}({\bf r},E)$.  
We use Lorenz function of half-width $0.001E_{\rm F}$ for $\delta(E)$ here. 
Left (right) panels are for the parallel (anti-parallel) vortex winding. 
}
\label{f3}
\end{figure}
%%%%%%%%%%%%%%%%%%%%%%%%%%%%% 

To study the differences due to vortex winding directions in other energy states, 
in Fig.~\ref{f3}(a), we present the $E$-dependence of $N({\bf r},E)$ and 
$e^{-i(w+1)\theta}{\cal F}_{s,{\rm triplet}}({\bf r},E)$ at $r \!=\! 2 k^{-1}_{\rm F}$. 
We find $e^{-i(w+1)\theta}{\cal F}_{s,{\rm triplet}}({\bf r},E) \!\ne\! N({\bf r},E)$ for parallel vortex winding, except for $E\!=\!0$. 
Only at $E\!=\! 0$, $e^{-i(w+1)\theta}{\cal F}_{s,{\rm triplet}}({\bf r},E)$ has the same peak structure as that of $N({\bf r},E)$, 
while it has negative value at $E \!\ne\! 0$. 
In the anti-parallel vortex winding case, we confirm 
$e^{-i(w+1)\theta}{\cal F}_{s,{\rm triplet}}({\bf r},E) \!\propto\! N({\bf r},E)$ for any bound states with $|E| \!<\! |\Delta|$. 
For the scattering state at $|E| \!>\! |\Delta|$, however, this relation does not hold. 
%${\cal F}_{s,{\rm triplet}}({\bf r},E) \!\ne\!  N({\bf r},E)$. 

%%%%%%%%%%%%%%%%%%%%%%%%%%%%% 
\begin{figure}[t!]
\includegraphics[width=8.0cm]{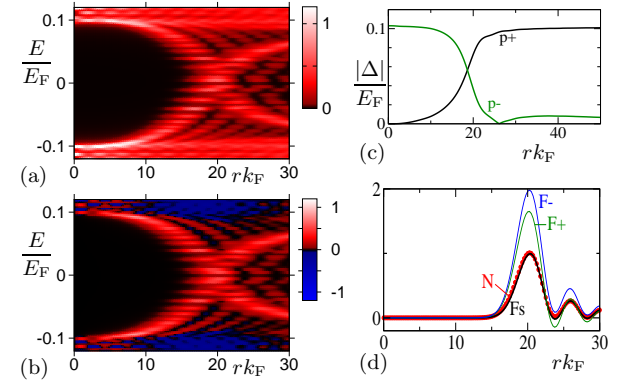}
\caption{
(Color online) 
Density plot of (a) $N({\bf r},E)$ and 
(b) ${\cal F}_{s,{\rm triplet}}({\bf r},E)$ 
for the double winding vortex with $w \!=\! -2$.
(c) $r$-dependence of $|\Delta_+({\bf r})|$ and  $|\Delta_-({\bf r})|$. 
(d)
$r$-dependence of LDOS $N({\bf r},E)$, 
${\cal F}_{s,{\rm triplet}}({\bf r},E)$, and 
${\cal F}_{\pm,{\rm triplet}}({\bf r},E)$ 
in the lowest energy state at 
$E=0.005E_{\rm F}$.  
We set $\theta \!=\! 0$. 
}
\label{f4}
\end{figure}
%%%%%%%%%%%%%%%%%%%%%%%%%%%%% 

The $E$-dependences of odd-frequency ${\cal F}_{s,{\rm triplet}}({\bf r},E)$ 
and even-frequency ${\cal F}_{\pm,{\rm triplet}}({\bf r},E)$ 
are compared in Fig.~\ref{f3}(b). 
From the symmetry relation for $E \!\leftrightarrow\! -E$, 
${\cal F}_{s,{\rm triplet}}$ (${\cal F}_{\pm,{\rm triplet}}$) is the even- (odd-) function of the real energy $E$. 
For anti-parallel vortex winding, we see that 
${\cal F}_{+,{\rm triplet}}({\bf r},E) \!\sim\! {\cal F}_{-,{\rm triplet}}({\bf r},E)$. 
However, this is not satisfied for parallel vortex winding. 
The odd-frequency pair potential 
$
\Delta_{s,{\rm triplet}}({\bf r})
\!\propto\! \int 
{\cal F}_{s,{\rm triplet}}({\bf r},E) f(E)
{\rm d}E 
$ 
has to vanish, as confirmed in the formulation on Matsubara frequencies. 
Thus, for ${\cal F}_{s,{\rm triplet}}({\bf r},E) $, 
positive contributions at low energy are canceled by 
the negative contribution at higher energies.

%%%%%%%%%%%%%%%%%%%%%%%%%%%% %%%%%%%%%%%%%%%%%%%%%%%%%%%% 
%\section{Double winding vortex} 

Lastly, we study the case of a double winding vortex. 
Here, zero-energy Majorana states do not appear, since the energy levels of the vortex bound state for $w \!=\!\pm2$ 
have a small gap of the order $\Delta^2/E_{\rm F}$.~\cite{TanakaSSC,Vinokur,MizushimaPRA2010} 
Therefore, this vortex does not exactly holds $e^{-i(w+1)\theta}{\cal F}_{s,{\rm triplet}}({\bf r},E) \!\ne\! N({\bf r},E)$ inside the vortex core.
However, in vortex bound states for $w \!=\! -2$, 
$N({\bf r},E)$ in Fig.~\ref{f4}(a) and 
${\cal F}_{s,{\rm triplet}}({\bf r},E)$ 
in Fig.~\ref{f4}(b) have similar structure each other. 
In the chiral $p$-wave pairing, 
since the induced $p_-$ component shares the core region without winding ($w+2\!=\!0$), 
there appears the chiral domain wall between outer $p_+$ and inner $p_-$ 
regions at $r \!\sim\! 19 k_{\rm F}^{-1}\!\sim\! \xi$ as 
in Fig.~\ref{f4}(c).~\cite{SaulsEschrig} 
Thus, while the low energy bound states for $w \!=\! -2$ appear far from the vortex center, 
they reduce to the bound states at the domain wall. 
The quasiparticle bound at the domain wall has similar structure 
to surface Majorana state in chiral $p$-wave superconductors.~\cite{MizushimaPRA2010} 
It is seen from Fig.~\ref{f4}(d) that the relations 
$e^{-i(w+1)\theta}{\cal F}_{s,{\rm triplet}}({\bf r},E) \!\propto\! N({\bf r},E)$ and 
${\cal F}_{+,{\rm triplet}}({\bf r},E) \!\sim\! {\cal F}_{-,{\rm triplet}}({\bf r},E)$ approximately hold in the low energy, while there are small deviations.
This is similar behavior to that of $w \!=\! -1$ in Fig.~\ref{f2}. 

The LDOS $N({\bf r},E)$, which we present in Figs.~1-4, is detectable through a STM-STS experiment. Since it is demonstrated in Figs.~2 and 3 and Eq.~(\ref{eq:Fodd}) that the odd-frequency pairing amplitude, $\mathcal{F}_{\rm s,triplet}({\bf r},E)$, traces the spatial shape of the zero energy LDOS, the characteristic behaviors may be unveiled by using superconducting STM.~\cite{Yokoyama} The local Josephson current between the superconductor and the superconducting STM tip is allowed only when the symmetry of local pair amplitudes in the superconductor is matched with that of the superconducting STM tip. Hence, the sharp peak of the odd-frequency $s$-wave pair amplitude around the vortex core is responsible to the local Josephson coupling with a superconducting STM tip with the same symmetry. For the anti-parallel vortex state, as shown in Fig.~3(a), the STM experiments with an odd-frequency $s$-wave superconducting tip may observe the spatial variation similar to the LDOS. As shown in Eq.~(\ref{eq:Feven}), however, the zero energy DOS does not contribute to even-frequency pair amplitudes, $\mathcal{F}_{\pm,{\rm triplet}}({\bf r},E\!=\!0)$, which is not responsible to a STM with an even-frequency chiral $p$-wave superconducting tip. This might potentially be a signature of the odd-frequency pairing.

Before closing, we discuss candidates materials of 
chiral $p$-wave superconductor and superfluid. 
One of the experimentally accessible superconductor is 
Sr$_{2}$RuO$_{4}$\cite{Maeno}
where surface Andreev bound state specific to chiral $p$-wave 
symmetry has been observed.\cite{Kashiwaya}
Recently, there have been proposed 
several heterostructures topologically 
equivalent to chiral $p$-wave 
superconductor, $e.g.$, 
topological insulator/spin-singlet $s$-wave superconductor heterostructures,\cite{Fu2008,Fu2009,Law} 
semiconductor/spin-singlet $s$-wave superconductor 
junctions with strong spin-orbit coupling.\cite{Sato,DasSarma2010,Alicea,Potter} 
The chiral superfluidity 
in Fermi gases near $p$-wave Feshbach resonances and spin-orbit coupled Fermi gases 
are also promising candidate systems.~\cite{MizushimaPRL2008,Inada,Zhai}
We hope our theoretical prediction will be experimentally verified in 
these systems. 

\section{Summary}

In summary, we have examined the relation of the odd-frequency $s$-wave pair amplitude 
${\cal F}_{s,{\rm triplet}}({\bf r},E)$ and LDOS $N({\bf r},E)$ 
for the quantum limit of CdGM vortex bound states in chiral $p$-wave superconductors, based on the BdG equation. 
The relation of Eq.~(\ref{N-OFs}) holds in the Majorana zero modes of a single winding vortex. Further, we have confirmed that ${\cal F}_{s,{\rm triplet}}({\bf r},E)$ and 
$N({\bf r},E)$ for the bound states 
have the same spatial structures even in finite $E$ 
at the core of the vortex with $w \!=\! -1$,   
and approximately at chiral domain wall of the $w \!=\! -2$ vortex. 
These spatial structures of the odd-frequency pair amplitude give valuable information, 
when we identify the odd-frequency pair amplitude in future experiments 
and discuss the associated anomalous interference phenomena.

\section*{ACKNOWLEDGMENTS}

%%%% 
We thank K. Machida for fruitful discussions. 
This work was supported by JSPS
(No.~2074023303) and 
``Topological Quantum Phenomena'' (No.~22103005) KAKENHI on innovation areas from MEXT.
% 2074023303: Mizushima (until Mar. 2012)
%This work was supported by KAKENHI 00000000. 


\begin{thebibliography}{9} 
\bibitem{Bergeret}
F.S. Bergeret, A.F. Volkov, and K.B. Efetov,  
Phys. Rev. Lett. {\bf 86}, 4096 (2001).

\bibitem{BergeretRMP}
F.S. Bergeret, A.F. Volkov, and K.B. Efetov, 
Rev. Mod. Phys. {\bf 77}, 1321 (2005).

\bibitem{Eschring}
M. Eschrig, T. L\"{o}fwander, T. Champel, J. C. Cuevas, J. Kopu and G. Sch\"{o}n;  
J. Low Temp. Phys. {\bf 147}, 457 (2007); 
M.\ Eschrig and T.\ L\"{o}fwander, Nature Phys.\ \textbf{4}, 138 (2008).

\bibitem{yokoyama}
T. Yokoyama, Y. Tanaka, and A. A. Golubov, Phys. Rev. B \textbf{75}, 094514 (2007). 


\bibitem{ya07sfs} 
Y.\ Asano, Y.\ Tanaka, and A.~A.\ Golubov, Phys.\ Rev.\ Lett.\ \textbf{98}, 107002 (2007); 
Y. Sawa, T. Yokoyama, Y. Tanaka, and A. A. Golubov, Phys. Rev. B \textbf{75}, 134508 (2007).

\bibitem{linder}
J.\ Linder, T.\ Yokoyama, A.\ Sudb{\o}, and M.\ Eschrig, Phys.\ Rev.\ Lett.\ \textbf{102}, 107008 (2009).


\bibitem{TanakaPRL2007v1}
Y. Tanaka and A. A. Golubov, Phys. Rev. Lett. {\bf 98}, 037003 (2007).

\bibitem{TanakaPRL2007}
Y. Tanaka, A.A. Golubov, S. Kashiwaya, and M. Ueda, 
Phys. Rev. Lett. {\bf 99}, 037005 (2007); 
Y.\ Tanaka, Y.\ Tanuma, A.~A.\ Golubov, 
Phys.\ Rev.~B \textbf{76}, 054522 (2007).

\bibitem{Asano07}
Y.\ Asano, Y.\ Tanaka, A.~A.\ Golubov, and S.\ Kashiwaya, Phys.\ Rev.\ Lett.\ \textbf{99}, 067005 (2007). 
 


\bibitem{Yokoyama}
T. Yokoyama, Y. Tanaka and A. A. Golubov, 
Phys. Rev. B. {\bf 78}, 012508 (2008). 

\bibitem{Yokoyama2}
T. Yokoyama, M. Ichioka, and Y. Tanaka, J. Phys. Soc. Jpn. {\bf 79}, 034702 (2010).

\bibitem{TanakaJPSJ2012}
Y. Tanaka, M. Sato, and N. Nagaosa, 
J. Phys. Soc. Jpn. {\bf 81}, 011013 (2012). 


\bibitem{TanakaReview}
Y. Tanaka and S. Kashiwaya, 
Phys. Rev. Lett. {\bf 74} 3451 (1995); 
S. Kashiwaya and Y. Tanaka, Rep. Prog. Phys. 
{\bf 63} 1641 (2000);  Y. Asano, Y. Tanaka, and S. Kashiwaya: Phys. Rev. B {\bf 69} (2004)
134501 (2004). 

\bibitem{Proximity}
Y.\ Tanaka and S.\ Kashiwaya, Phys.\ Rev.~B \textbf{70}, 012507 (2004); Y. Tanaka, S. Kashiwaya, and T. Yokoyama, ibid \textbf{71}, 094513 
(2005); Y.\ Tanaka, Y.\ Asano, A.~A.\ Golubov, and S.\ Kashiwaya, Phys.\ Rev.~B \textbf{72}, 140503(R) (2005);  
Y.\ Asano, Y.\ Tanaka, and S.\ Kashiwaya, Phys.\ Rev.\ Lett.\ \textbf{96}, 097007 (2006); T.\ Yokoyama, Y.\ Tanaka, and N.\ Nagaosa,
Phys. Rev. Lett. \textbf{106}, 246601 (2011); 
Y. Tanaka, Y. Nazarov, and S. Kashiwaya: Phys. Rev. Lett. \textbf{90} 
167003 (2003).

\bibitem{Berezinskii}
V.L. Berezinskii: 
JETP Lett. {\bf 20} (1974) 287.

\bibitem{Balatsky} 
A.\ Balatsky and E.\ Abrahams, Phys.\ Rev.~B \textbf{45}, 13125 (1992).

\bibitem{Caroli} 
C. Caroli, P.G. de Gennes, and J. Matricon,  
Phys. Lett. {\bf 9}, 307 (1964).

\bibitem{Hayashi}
N. Hayashi, T. Isoshima, M. Ichioka, and K. Machida, 
Phys. Rev. Lett. {\bf 80}, 2921 (1998).

\bibitem{kaneko}
S. Kaneko, M. Hafiz, E. Kakizaki, K. Matsuba, N, Nishida, T. Kawakami, T. Mizushima, and K. Machida, 
J. Phys. Soc. Jpn. {\bf 81}, 063701 (2012).

\bibitem{MIT}
M. W. Zwierlein, J. R. Abo-Shaeer, A. Schirotzek, C. H. Schunck, and W. Ketterle, 
Nature (London) {\bf 435}, 1047 (2005).

\bibitem{ReadGreen}
N. Read and D. Green,  
Phys. Rev. B {\bf 61}, 10267 (2000). 

\bibitem{ivanov}
D. A. Ivanov, 
Phys. Rev. Lett. {\bf 86}, 268 (2001).

\bibitem{MizushimaPRL2008}
T. Mizushima, M. Ichioka, and K. Machida,  
Phys. Rev. Lett. {\bf 101}, 150409 (2008). 

\bibitem{MizushimaPRA2010}
T. Mizushima and K. Machida, 
Phys. Rev. A {\bf 81}, 053605 (2010). 

\bibitem{tewari}
S. Tewari, S. Das Sarma, and D.-H. Lee, 
Phys. Rev. Lett. {\bf 99}, 037001 (2007).

\bibitem{gurarie}
V. Gurarie and L. Radzihovsky, Phys. Rev. B {\bf 75}, 212509 (2007).

\bibitem{kawakami}
T. Kawakami, T. Mizushima, and K. Machida, J. Phys. Soc. Jpn. {\bf 80}, 044603 (2011).


\bibitem{yasui}
S. Yasui, K. Itakura, and M. Nitta, Nucl. Phys. B {\bf 859}, 261 (2012).

\bibitem{Higashitani}
S. Higashitani, S. Matsuo, Y. Nagato, and K. Nagai, 
S. Murakawa, R.Nomura, and Y. Okuda,  
Phys. Rev. B {\bf 85}, 024524 (2012); 
S. Higashitani, Y. Nagato and K. Nagai, J. Low Temp. Phys. 
{\bf 155}, 83 (2009).

\bibitem{asano2012}
Y. Asano and Y. Tanaka, arXiv:1204.4226.

\bibitem{Matsumoto}
M. Matsumoto and R. Heeb, Phys. Rev. B {\bf 65}, 014504 (2001).

\bibitem{kato1}
Y. Kato, J. Phys. Soc. Jpn. {\bf 69}, 3378 (2000). 

\bibitem{kato2}
Y. Kato and N. Hayashi, 
J. Phys. Soc. Jpn. {\bf 70}, 3368 (2001); {\it ibid} {\bf 71}, 1721 (2002).


\bibitem{tanuma}
Y. Tanuma, N. Hayashi, Y. Tanaka, and A. A. Golubov, 
Phys. Rev. Lett. {\bf 102}, 117003 (2009).

\bibitem{Gygi}
F. Gygi and M. Schl\"{u}ter, Phys. Rev. B \textbf{43}, 7609 (1991). 
%Y. Tanaka, A. Hasegawa and H. Takayanagi, Solid State Commun. 
%{\bf 85}, 321 (1993). 
 
\bibitem{TanakaSSC}
Y. Tanaka, H. Takayanagi, and A. Hasegawa, 
Solid State Commun. {\bf 85}, 321 (1993).

\bibitem{memo}
The symmetric relation for $E\!\leftrightarrow\!-E$ holds in the imaginary part of the Green's function $e^{-i(w+1)\theta}\mathcal{F}_s$, while the real part is anti-symmetric. See for example, Y. Tanaka, Y. Asano, A. A. Golubov, and S. Kashiwaya, Phys. Rev. B {\bf 72}, 140503(R) (2005); Y. Tanaka and S. Kashiwaya, Phys. Rev. B {\bf 70}, 012507 (2004) and Ref.~\onlinecite{TanakaPRL2007}.

\bibitem{Vinokur}
A. S. Mel'nikov and V. M. Vinokur, 
Nature (London) {\bf 415}, 60 (2002).

\bibitem{SaulsEschrig}
J.A. Sauls and M. Eschrig, 
New J. Phys. {\bf 11}, 075008 (2009).




\bibitem{Maeno}
A. P. Mackenzie and Y. Maeno, Rev. Mod. Phys. \textbf{75}, 657
(2003).


\bibitem{Kashiwaya}
 S. Kashiwaya, H. Kashiwaya, H. Kambara, T. Furuta,
H. Yaguchi, Y. Tanaka, and Y. Maeno, Phys. Rev. Lett.
\textbf{107}, 077003 (2011). 

%%%%%%%%%%%%%%%%%%%%%%%%%%%%%%%%%%%%%%%%%%%%%%%%%%%%%
% Majorana from topological classification
%%%%%%%%%%%%%%%%%%%%%%%%%%%%%%%%%%%%%%%%%%%%%%%%%%%%%%%

\bibitem{Fu2008}
L. Fu and C. L. Kane, Phys. Rev. Lett. \textbf{100}, 096407 
(2008). 

%%%%%%%%%%%%%%%%%%%%%%%%%%%%%%%%%%%%%%%%%%%%%%%%%%%%%%%%%%%%%%%%%%%%%%%%%%%%%%%%% Topological insulator and Majorana
%%%%%%%%%%%%%%%%%%%%%%%%%%%%%%%%%%%%%%%%%%%%%%%%%%%%%%%%%%%
\bibitem{Fu2009}
L. Fu and C. L. Kane, Phys. Rev. Lett. \textbf{102}, 
216403 (2009); A. R. Akhmerov, J. Nilsson, and C. W. J. Beenakker,
Phys. Rev. Lett. \textbf{102}, 216404 (2009); 
Y. Tanaka, T. Yokoyama, and N. Nagaosa, Phys. Rev.
Lett. \textbf{103}, 107002 (2009); 
J. Linder, Y. Tanaka, T. Yokoyama, A. Sudbo, and
N. Nagaosa, Phys. Rev. Lett. \textbf{104}, 067001 (2010). 

\bibitem{Law}
K. T. Law, P. A. Lee, and T. K. Ng, Phys. Rev. Lett.
\textbf{103}, 237001 (2009).


\bibitem{Sato}
M. Sato, Y. Takahashi, S. Fujimoto, Phys. Rev. 
Lett \textbf{103}, 020401 (2009); 
M. Sato, Y. Takahashi, and S. Fujimoto, Phys. Rev. B 
\textbf{82}, 134521 (2010).


\bibitem{DasSarma2010}
J.D. Sau, R.M. Lutchyn, S. Tewari, and S. Das Sarma,
Phys. Rev. Lett. \textbf{104}, 040502 (2010); 
T. D. Stanescu, J. D. Sau, R. M. Lutchyn, and S. Das Sarma,
Phys. Rev. B, \textbf{81}, 241310 (2010). 

\bibitem{Alicea}
J. Alicea, Phys. Rev. B \textbf{81}, 125318 (2010). 

\bibitem{Potter}
A. C. Potter and P. A. Lee, 
Phys. Rev. Lett., \textbf{105}, 227003 (2010); 
J. Linder and A. Sudbo, 
Phys. Rev. B \textbf{82}, 085314 (2010); 
A. Yamakage, Y. Tanaka and N. Nagaosa, 
Phys. Rev. Lett. \textbf{108}, 087003 (2012). 

\bibitem{Inada}
Y. Inada, M. Horikoshi, S. Nakajima, M. Kuwata-Gonokami, M. Ueda, and T. Mukaiyama,
Phys. Rev. Lett. {\bf 101}, 100401 (2008) and references therein.

\bibitem{Zhai}
P. Wang, Z.-Q. Yu, Z. Fu, J. Miao, L. Huang, S. Chai, H. Zhai, and J. Zhang, 
arXiv:1204.1887.

\end{thebibliography}
\end{document}